\documentclass{PoS}

\newcommand\ital{\emph}
\newcommand\doi[1]{\href{http://dx.doi.org/#1}{[doi:#1]}}
\newcommand\eprint[1]{\href{http://arXiv.org/abs/#1}{[arXiv:#1]}}

\PoS{PoS(LAT2005)136}

\title{Towards a chiral gauge theory by deconstruction in AdS5}

\ShortTitle{Towards a chiral gauge theory by deconstruction in $AdS_5$}

\author{\speaker{Tanmoy Bhattacharya}, Rajan Gupta, Matthew R. Martin,
        and Yuri Shirman \\
        Los Alamos National Laboratory\\
        E-mail: \email{tanmoy@lanl.gov},
                \email{rajan@lanl.gov},
                \email{mrmartin@lanl.gov},
                \email{shirman@lanl.gov}}
\author{Csaba Cs\'aki\\
        Cornell University\\
        E-mail: \email{csaki@lepp.cornell.edu}}
\author{John Terning\\
        University of California, Davis\\
        E-mail: \email{terning@physics.ucdavis.edu}}

\abstract{We describe an implementation of a deconstructed gauge
theory with charged fermions defined on an interval in five
dimensional AdS space. The four dimensional slices are Minkowski, and
the end slices support four dimensional chiral zero modes. In such a
theory, the energy scales warp down as we move along the fifth
dimension. If we augment this theory with localized neutral
4-dimensional Majorana fermions on the low energy end, and implement a
Higgs mechanism there, we can arrange the theory such that the
lightest gauge boson mode and the chiral mode on the wall at the high
energy end are parametrically lighter than all the other states in the
theory.  If this semiclassical construction does not run into problems
at the quantum level, this may provide an explicit construction of a
chiral gauge theory.  Instanton effects are expected to make the gauge
boson heavy only if the resulting effective theory is anomalous.

}

\FullConference{XXIIIrd International Symposium on Lattice Field Theory\\
                 25-30 July 2005\\
                 Trinity College, Dublin, Ireland}

\ifx\pdfstartlink\undefined\let\pdfstartlink\pdfannotlink\fi
\begin{document}

\section{Introduction}

Chiral gauge theory of weak interactions forms an important ingredient
of the standard model, but so far chiral gauge theories have defied a
definition beyond perturbation theory.  These in fact form the only
class of perturbatively defined field theories with no
non-perturbative formulation. This problem is gaining practical import
since a number of proposed extensions to the standard model invoke
strongly interacting chiral dynamics, and a quantitative analysis
requires the evaluation of non-perturbative condensates in these
theories. The only known non-perturbative regulator in four-dimensions
is the lattice formulation and we investigate it here.

A fundamental problem in realizing a chiral gauge theory on the
lattice is encapsulated in the Nielsen Ninomiya
theorem~\cite{NielsenNinomiya}.  According to this theorem, under mild
conditions of locality and analyticity of the propagator, a
translationally invariant fermion theory that preserves the continuum
chiral symmetry exactly on the lattice has paired left and right
chiral modes in the continuum limit.  Since these modes are related by
boosts in the discretized theory, they cannot be differently charged
under any symmetry preserved by the discretization, and any na\"\i{}ve
attempt at obtaining an unbroken chiral gauge theory must fail.  This
result is also expected based on our knowledge that chiral gauge
theories are undefined if anomalies do not cancel. The anomaly
cancellations can occur, as in the standard model, between fermion
species that interact only through the gauge fields. This cancellation
is difficult to preserve in any na\"\i{}ve discretization.

A way out of this conundrum was suggested by Kaplan~\cite{Kaplan} by
extending fermionic gauge theories to five dimensions.  Theories on a
five dimensional interval typically have \emph{edge} states that are
chiral under the four dimensional Euclidean group acting on the edge
of the region.  These states still come in chiral pairs, 
but one can arrange the parameters to locate the left
and right chiral modes on the different edges, and hence separate them in
the fifth dimension.  Since the fifth dimension is unphysical, one is
free to add interactions that are not translationally invariant along
this direction.  Previous attempts to realize a chiral gauge theory by
confining the gauge interactions to only a part of the five
dimensional space were, however, unsuccessful.~\cite{waveguide}.

In this paper we consider an alternate formulation with a truly five
dimensional gauge field~\cite{us}.  The translational invariance of this gauge
interaction is broken by invoking a Higgs' mechanism using extra
matter fields propagating only along one edge.  This Higgs' phenomenon
makes the fermion states at that edge heavy, but has negligible effect
on states located elsewhere in the five dimensional world.  A
classical analysis reveals that in the presence of background
five-dimensional space-time curvature we can take a limit in which a
four-dimensional gauge boson and one chiral fermion remain massless,
whereas all the other states in the theory become infinitely heavy and
decouple. If the resulting gauge theory has an uncancelled triangle
anomaly then this construction of a chiral gauge theory is easily seen
to fail at the quantum level. A complete quantum analysis remains
beyond our reach because even perturbation theory is not valid in some
regions in this five dimensional space.

\section{The flat space analogue}
\label{sec:flat}

\begin{figure}
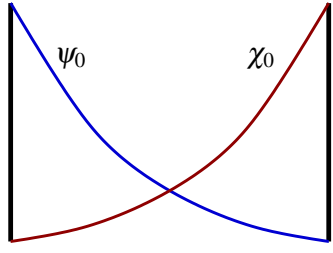

\begin{center}
\leavevmode\vbox{\hbox{%
\includegraphics[width=2in]{profile-0.mps}%
\llap{\includegraphics[width=2in]{profile-1.mps}}%
\llap{\includegraphics[width=2in]{profile-2.mps}}%
\llap{\vbox to 1.1in{\hbox to 1.6in{%
\color{black}$\psi_0$\hspace*{0.8in} $\chi_0$\hss}\vss}}}
}
\end{center}
\caption{The exponentially localized edge states of the fermions,
  \(\psi_0 \propto \exp{-Mz}\) and \(\bar\chi_0 \propto \exp{Mz}\).}
\label{fig:ferm}
\end{figure}

Before constructing the full model, it is instructive to study an
analogue construction in flat five-dimensional space.  We consider 
this world bounded by flat Euclidean slices separated by distance
\(R'-R\).  A free massive fermion in this spacetime is described
by the action
\begin{eqnarray}
  S &=& \int d^4\,x \int_R^{R'} d\,z \big\{ 
   -i\,\bar\psi\partial_\mu\bar\sigma^\mu\psi
   -i\,\chi\partial_\mu\sigma^\mu\bar\chi
   +\psi\partial_z\chi-\bar\chi\partial_z\bar\psi \nonumber\\
    &&\qquad\qquad\qquad\qquad {} + M\,\psi\chi + M\,\bar\chi\bar\psi \big\}\,,
     \nonumber
\end{eqnarray}
where \(\psi\) and \(\chi\) are two-component fermion fields of opposite
chirality, \(\sigma^\mu\) are the Pauli matrices (or identity for the
time direction), \(M\) is the five dimensional mass, and the overbar
represents conjugation.  A Kaluza-Klein decomposition gives modes
\(\chi_n\) and \(\psi_n\) that satisfy
\[
   (\partial_z - M)\bar\chi_n = m_n\bar\chi_n \qquad\qquad\qquad
  -(\partial_z + M)\psi_n = m_n\psi_n\,.
\]
The edge states are the states with \(m_0=0\) and are exponentially
localized at the left and right walls (Fig.~\ref{fig:ferm}). The
mass scaleof the rest of the states in the theory, \(\pi/2(R'-R)\), 
is set by the length of the fifth dimension.

A gauge theory is obtained by changing the derivatives
above to gauge covariant ones and adding a kinetic term for the gauge
bosons
\[
    \int d^4\,x \int_R^{R'} d\,z \frac1{4g_5^2}
      \left( F_{\mu\nu} F^{\mu\nu} + 2 F_{\mu5}F^{\mu5} \right)\,.
\]
In such a theory the left and right chiral modes, \(\psi_0\) and
\(\chi_0\), have equal charge, and one needs to decouple one of them
(by making it heavy), say \(\chi_0\) on the right wall, to obtain a
chiral gauge theory.  To this end, we introduce an Higgs' mechanism
localized on the right wall, and use the Higgs' field \(H\) to mix
\(\chi\) with a neutral fermions \(S_L\) with large Majorana mass,
\(m\):\footnote{The addition of a neutral right handed fermion \(S_R\)
at \(R'\) along with its Yukawa coupling to the bulk field
\(\bar\psi\) has, however, very little impact on the zero mode
\(\psi_0\) exponentially localized at \(R\).  Such an addition is
necessary in the actual construction but is dropped in this exposition
for brevity.}
\[
  y\,\chi H S_L + m S_L S_L + \mbox{h.c.}\,
\]
where \(y\) is the Yukawa coupling.  This provides a mass of order \(y
\langle H\rangle\) to \(\chi_0\). Since Yukawa couplings are
infrared free it is difficult to make \(y\) much larger than the
gauge coupling \(g\).  This is problematic since for small values of
\(\langle H\rangle\) the gauge boson also acquires a mass of order
\(g \langle H\rangle\), and it appears one cannot take a limit where
the unwanted mode \(\chi_0\) decouples, but the gauge boson stays in
the spectrum.

The situation is qualitatively different when the vacuum expectation
value is large.  The odd Kaluza-Klein modes of the gauge boson, which
vanish at the boundary, do not feel the effect of the Higgs' mechanism
at the classical level and retain a mass, \({\tilde m}_1\), of the order of \(\pi/2(R'-R)\). 
The even Kaluza-Klein modes pick up large masses by this Higgs'
mechanism.   So, if the length of the fifth dimension is large, we do
find light gauge bosons and a massless chiral mode, \(\psi_0\) in the
spectrum\footnote{The masses of the Kaluza-Klein modes \(\chi_n\),
controlled by the scale \(M\) and not by the size of the extra
dimension, stay heavy.} as required to construct a chiral gauge
theory. 

This construction, however, fails since \emph{all and not only the
lightest} of the odd Kaluza-Klein modes of the gauge field are
controlled by the same scale \(\pi/2(R'-R)\).  Thus in the limit that
the extra dimension becomes large, the theory reverts to being five
dimensional.

\section{The Model in \(AdS_5\)}
\label{sec:model}

It has been known for a while~\cite{pheno} that if we consider an
interval in a curved five dimensional space, the gravitational
acceleration towards one wall can give rise to an effective four
dimensional theory \emph{even when the length of the extra dimension
is large.}  Accordingly, we consider a Euclidean version of a slice of
\(AdS_5\) bounded by four-dimensional flat Minkowski slices separated
by a proper distance denoted by \(\ln R'/R\). \(AdS_5\) is a
homogeneous five-dimensional space with a constant negative radius of
curvature, which is also denoted by \(-R\). The comparison with our
flat space analysis is easiest in the metric
\[
  ds^2 = \left(\frac Rz \right)^2 
      \left(\eta_{\mu\nu} dx^\mu dx^\nu - dz^2\right)\,,
\]
with the chosen interval being \((R,R')\). In this background, the
odd Kaluza-Klein modes of the gauge boson have masses given by
\[
  {\tilde m}_1^2 = O\left(\frac1{R'^2\ln\left(R'/R\right)}\right)\,,\qquad\qquad\qquad
  {\tilde m}_n^2 \approx O\left(\frac1{{R'}^2}\right)\,,
\]
whereas the fermion modes are not significantly affected for large
\(M\).  This lets us take the limit
\[
   R'\to 0\, \qquad\qquad\qquad\qquad\qquad R'^2 \ln (R'/R)\to\infty\,
\]
which decouples all the Kaluza-Klein modes leaving behind a massless
gauge boson interacting with the chiral fermion \(\psi_0\).

\section{Deconstruction}
\label{sec:deconstruction}

Five dimensional theories may not be renormalizable, and it may not
be possible to interpret the Lagrangian parameters as running
couplings at a certain scale. We, therefore, need to be careful in drawing
conclusions based on the tree level constructions provided in the
previous sections.  To study this systematically we choose to
deconstruct~\cite{deconstruction} the five dimensional theory into a
stack of four dimensional continuum slices placed at discrete
positions along the fifth dimension.  To maintain as many of the space time
symmetries of the \(AdS_5\) space as possible, we choose the positions of the
slices to be 
\[
   z_i = \left(1 + a\right)^{i-1} R, \qquad i = 1\ldots N,\qquad 
    a = \exp\left(\frac1{N-1}\ln \frac R{R'}\right)-1
\]
where we refer to the small dimensionless number \(a\) as the ``lattice
spacing''. The modes at each of these four dimensional slices
can be reinterpreted as those of fields transforming under a separate four
dimensional gauge group.  It is easy to check that the gauge coupling
of the \(i^{\mbox{th}}\) gauge group is given by
\[
   \frac1{g_i^2} = \frac{aR}{g_5(z_i)^2}\,,
\]
where we have generalized the model to allow for a variation of
\(g_5\) along the fifth dimension.

The link fields in the fifth dimension, \({\cal P}
\exp\int_{z_i}^{z_{i+1}} i\,A_5\,dz\) can then be interpreted as
bifundamental scalars (transforming according to the fundamental
(anti-fundamental) representation under the gauge theory at \(z_i\)
(\(z_{i+1}\))).  With this identification, the original theory, before
we add the Higgs mechanism at \(R'\), can be seen to be invariant
under the product of the gauge groups at each site but realized in the
Higgs phase with the bifundamentals acquiring a vacuum expectation
value related to the lattice spacing.  This view of the
five-dimensional theory is called deconstruction.  The lowest gauge
boson mode is, not surprisingly, the discretization of the lowest
Kaluza-Klein mode: the equal superposition of the gauge fields on 
all the slices, and its effective coupling is given by
\[
  \frac1{g_4^2} \approx \sum_{i=1}^N \frac1{g_i^2}\,.
\]

The continuum \(AdS_5\) symmetry requires all lengths to scale in
proportion to the \(z\) coordinate as we move in the fifth dimension.
To maintain a discrete version of this we can choose the renormalized
couplings to satisfy
\[
  g_i^R(\mu_i) = constant \qquad\qquad\mbox{if}\qquad\qquad 
  \mu_i \propto 1/z_i\,.
\]
A one loop calculation, assuming $\mu \ll 1/R'$, then yields
\[
\frac1{g_4^2(\mu)} \approx \frac N{g_1^2(1/aR)} + 
                           \frac {\beta_0}{8\pi^2}\ln aR\mu\,
\]
whence we can obtain the \(\Lambda\) parameter of the theory associated 
with the lowest mode to be 
\[
  aR\Lambda = \exp \frac{-8\pi^2N}{\beta_0 g_1^2(1/aR)}\,.
\]

To obtain a chiral gauge theory we need to take the limit 
\[
   \frac{m_{KK}}\Lambda \to \infty\,, \qquad\qquad\qquad\qquad
   \frac{m_1}\Lambda \to 0\,,
\]
where \(m_{KK}\) is the typical Kaluza-Klein state and \(m_1\) is the
lightest gauge mode.  Using
\[
   m_{KK}^2 R^2 \sim (1 + a)^{2N}\,, \qquad\qquad\qquad\qquad
   \frac {m_1^2}{m_{KK}^2} \sim \frac1{N\log(1+a)}\,,
\]
one finds that the limit can be realized with the choice
\[
   N\to\infty \qquad\qquad\qquad\qquad 
   g_1^2(\frac1{aR}) \sim \frac{8\pi^2}{\beta_0 a}\,,
\]
with $a$ held fixed.

The arguments used in deriving these results, however, relied on the
wavefunctions on the lattice being a discretization of the continuum
tree-level wave functions.  This property holds only if the lattice
spacing \(a\) is much smaller than unity, and this in turn implies we
need to choose the lattice couplings to be large. As a result, the
one-loop results presented here can only be treated as
suggestive.\footnote{A more detailed argument shows that it is
indeed the coupling at the scale \(1/aR\) which controls the validity
of the required one-loop calculation.}

\section{Potential problem}
\label{sec:problem}

It is well known that the massless limit of a vector field theory is
singular, and often involves strong couplings.  It is instructive to
note that even in our construction, the `longitudinal' mode of the
gauge field becomes strongly interacting with an effective Yukawa like
interaction with the fermions. This interaction, of strength
\[
   y(z) \propto \frac {z\ln(z/R)} {R'\ln(R'/R)}\,,
\]
pushes the unwanted chirality fermion away from the wall with the
Higgs' mechanism, and can potentially make it light.  Explicit
evaluation in one-loop perturbation theory, however, shows that the
fermion decouples even in the presence of this interaction.
Unfortunately, the large gauge coupling precludes a pertubative
resolution of this question. To settle this issue a non-perturbative
calculation is required.

\end{document}